\documentclass[%
reprint,
 amsmath,amssymb,
 aps,
]{revtex4-1}

\usepackage{graphicx}
\usepackage{color}
\usepackage{dcolumn}
\usepackage{bm}

\usepackage{amsmath}
\DeclareFontFamily{U}{euc}{}
\DeclareFontShape{U}{euc}{m}{n}{<-6>eurm5<6-8>eurm7<8->eurm10}{}%
\DeclareSymbolFont{AMSc}{U}{euc}{m}{n}
\DeclareMathSymbol{\umu}{\mathord}{AMSc}{"16} 

\begin{document}

\title{An 8-bit carry look-ahead adder with 150\,ps latency and \\sub-microwatt power dissipation at 10\,GHz}
\author{Anna Y. Herr, Quentin P. Herr, Oliver T. Oberg, Ofer Naaman, John X. Przybysz, Pavel Borodulin, and Steven B. Shauck}
\affiliation{Northrop Grumman Systems Corp., Baltimore, Maryland 21240, USA}
\email[email: ]{anna.herr@ngc.com}
\thanks{\\This work was supported in part by the Defense Microelectronics Activity under the Advanced Technology Support Program.}



\begin{abstract}

Reciprocal Quantum Logic combines the speed and power-efficiency of
single-flux quantum superconductor devices with design features that
are similar to CMOS. We have demonstrated an 8-bit carry look-ahead
adder in the technology using combinational gates with fanout of four
and non-local interconnect. Measured power dissipation of the fully
active circuit is only 510\,nW at 6.2\,GHz. Latency is only 150\,ps at
a clock rate of 10\,GHz.

\end{abstract}

\maketitle

\section{Introduction}

Superconducting Reciprocal Quantum Logic (RQL) is an ultra-low-power
technology for high-performance computing that gives unmatched
efficiency in terms of operations per joule \cite{RQL11}. Simple
experiments have shown bit energy approaching 1000\,$k_BT$, with
further reduction expected using smaller devices. This means that the
technology offers two orders of magnitude power savings over 22\,nm
CMOS even after taking into account the overhead of the cryocooler, of
order 1000\,W/W at an operating temperature of 4.2\,K
\cite{Coolers}. RQL introduces a combination of new features that are
unattainable by other superconducting logic families
\cite{Likharev_91}, \cite{e-RSFQ}, including zero static power
dissipation, stable timing, low bit-error-rate, low active-device
count, and low-latency. The logic is combinational with multiple
levels of logic per pipeline stage, following conventional CMOS
behavior. This allows a wealth of CMOS design tools and methods to be
applied. A small-current AC waveform provides power and a stable clock
reference to each gate, which in principle allows the technology to
scale to VLSI.

Here we report the design, fabrication, and test of an RQL 8-bit carry
look-ahead (CLA) adder as a benchmark to demonstrate the effectiveness
of the logic in a larger circuit. The CLA adder is an important
hardware component for low-latency parallel addition. However,
implementation of this circuit has been a major challenge for
superconducting technology in the past. Gate-level pipelining and
inefficient clock distribution resulted in designs with high latency
that defeated the purpose \cite{bunyk_99}. Asynchronous or self-timed
designs can reduce latency \cite{kameda1999self, Filippov11}, but high
active device count, high current bias, and timing uncertainty limit
increased integration scale. The RQL implementation of the adder
achieves a tenfold improvement in latency and Josephson-junction
device count over earlier designs and shows that the main metric,
power efficiency, is scalable to a large circuit as predicted by the
first simple tests.

\section{Logical Design}

\begin{figure}[tbp]
\includegraphics{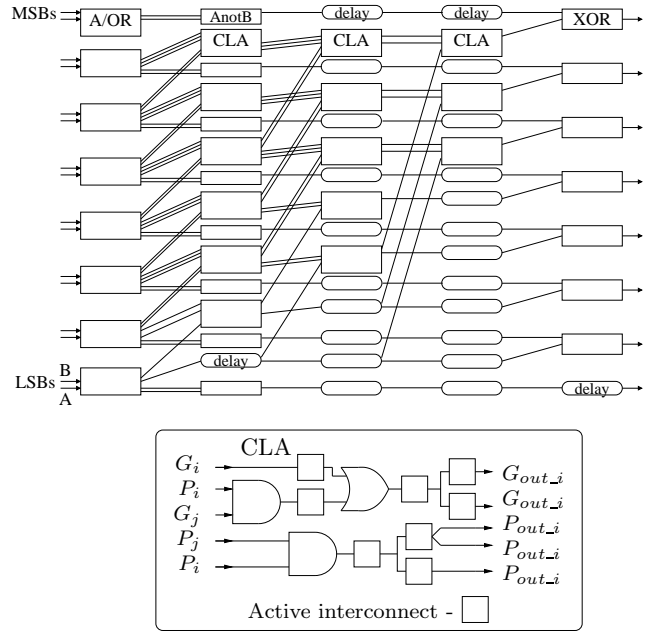}
\caption{Block diagram of the 8-bit Kogge-Stone CLA adder circuit and
  schematic of the individual CLA bit.
\label{fig1}}
\end{figure}

The adder, shown schematically in Fig.~\ref{fig1}, computes the carry
bits with minimal latency at the expense of four-way
fan-out and high-density, long interconnects. The design is a
Kogge-Stone radix-2 implementation that for N-bit inputs consists of
$\log_2N+2$ stages. The first stage (A/OR) produces carry propagate
and carry generate signals. The following stages form the carry
look-ahead (CLA) network that computes all carries in parallel. The
final result is computed in the last stage where carries are summed
with the input sums using logical XOR.

The A/OR stage produces logical AND and OR for input bits $A_i$ and
$B_i$, expressed as $G_i = A_iB_i$, and $P_i = A_i+B_i$. These outputs
are the generate and propagate signals, which have fanout of up to
four to reach multiple CLA blocks in the second stage. Each stage of
the CLA network takes select inputs from the previous stage $(G_i,
P_i)$ and $(G_j, P_j)$, and computes $(G_{out\_i}, P_{out\_i})$
expressed as $G_{out\_i} = P_iG_j+G_i$, and $P_{out\_i} = P_iP_j$.
For the $n^{th}$ stage, input selection obeys $j = i - 2^n$. Inputs
and outputs of the CLA blocks are pruned to stay within boundaries and
to produce only the culminating generate bits, which correspond to
carries $C_i$. The final sum is computed as $S_i = A_i \otimes B_i
\otimes C_i$.

RQL allows efficient design of the adder in terms of device count and
latency. Unlike other superconductor logic, RQL gates are
combinational with behavioral descriptions similar to CMOS. As in CMOS
circuits, this allows use of multiple levels of logic per pipe-line
stage that greatly reduces latency. RQL logic has an additional
efficiency in wave pipelining that eliminates active latches to
synchronize signals at each stage. Instead, four-phase power is used
to move signals from stage to stage on the rising edge of the power
waveform, which also serves as the clock. Data propagation speed finds
equilibrium independently of initial conditions, clock amplitude,
parameter variations, and data pattern. For the 8-bit adder, 8
parallel clock lines provide power to each bit. There is no skew
between clock lines provided that their geometrical lengths are equal.
Each of the five stages of the adder takes one phase of the clock,
which amounts to a total of 1.25 clock cycles for the computation to
complete.

RQL gates have no static power dissipation. Power is only dissipated
for logical ``ones'', physically encoded as a pair of positive and
negative (reciprocal) single-flux-quantum (SFQ) pulses. The dynamic
power dissipation is $P=0.33\,I_C \Phi_0 Nf$, where $I_C$ is the
average critical current of the Josephson junctions, $N$ is the number
of junctions, $\Phi_0=h/2e=2.068$\,mV\,ps is the flux quantum, $f$ is
operating frequency, and the prefactor of 0.33 is experimentally
determined \cite{RQL11}. AC loss in the superconducting line is small
\cite{loss}. The AC power is applied through weak inductive coupling,
which leads to important advantages: the gates are powered in series
requiring small current, and dynamics in the circuit do not affect
timing. An RQL circuit scaled to 2 million Josephson devices of
0.1\,mA average critical current would dissipate only 1.4\,mW of power
when fully active. Applied power would need to be larger than
dissipated power in order to maintain clock stability in terms of
amplitude and timing. Since operating margins of the gates are
sufficient to tolerate $\pm 10$\% variation in clock current, only
4\,mW of clock power would be needed to power the 2-million-device
chip, amounting to only 9\,mA of current on a 50\,$\Omega$
line. Timing variation would amount to only 5\,ps, or $\pm 2$\% of the
clock period at 10\,GHz \cite{RQL11}.

The adder was designed using gate-level VHDL models. Only two basic
RQL logic gates were used: AndOr and AnotB \cite{RQL11}. The AndOr
gate propagates the first logical ``one'' input in a given clock cycle
to the Or output, and the second ``one'' input to the And output.  The
AnotB gate propagates a logical ``one'' on the A input to the output
unless a logical ``one'' on the B input comes in the same clock cycle.
Logical XOR can be expressed as ``A or B, but not both A and B,''
which is implemented by connecting the outputs of the AndOr gate to
the inputs of AnotB gate. The CLA block, shown in Fig.~\ref{fig1},
contains And and Or gates. These are implemented by pruning the
outputs of the AndOr gate. RQL gates require active interconnect,
consisting of two sequential Josephson junctions, for signal
amplification. The same active interconnect is used both in splitters
to produce fanout, and to produce a delay element. The five
less-significant carries use delay cells to reach the last column. The
AnotB gates in the second column produce the partial sum are in
parallel with the CLA network and also use delay cells. The three
more-significant carries set the overall latency of the circuit, and
achieve up to eight levels of logic for a pipeline stage consisting of
all four clock phases.

\begin{figure}[tbp]
\includegraphics{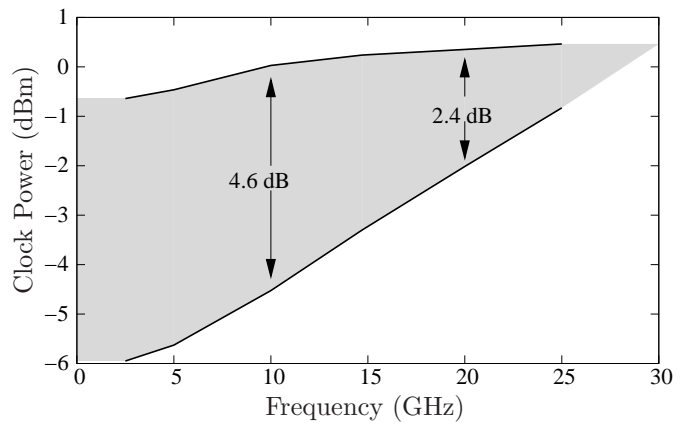}
\caption{Simulated operating margins on clock power supplied to the
  adder, as a function of clock rate.
\label{fig2}}
\end{figure}

The adder design was verified using WRSpice, a physical-level circuit
simulator that includes a device model for the Josephson junction
\cite{wrspice}. Simulated operating margins on clock power as a
function of frequency for the entire eight-bit adder are shown in
Fig.~\ref{fig2}. The upper limit corresponds to over-bias and is
relatively frequency independent. The lower limit narrows the
operating region with frequency, indicating that the circuit is
latency-limited at high clock rate. The Josephson junction model used
was for 1.5\,$\umu$m devices with a critical current density of
4.5\,kA/cm$^2$ and an $I_cR$ product of 0.7\,mV, where $I_c$ is the
critical current, typically 141-200\,$\umu$A in our circuit, and $R$ is
the external shunt resistance. Such devices produce voltage pulses
equal to the single-flux quantum, about 0.7\,mV high and 3\,ps
wide. Device delay for sequentially wired junctions is also about
3\,ps under nominal clock power.  A clock rate of 10\,GHz was the
design target as there are up to eight sequential junctions per phase
and four phases per clock cycle. In the circuit simulation, this clock
rate gives a wide clock-power operating margin of 4.6\,dB.

\section{Physical design}

\begin{figure}[tbp]
\includegraphics{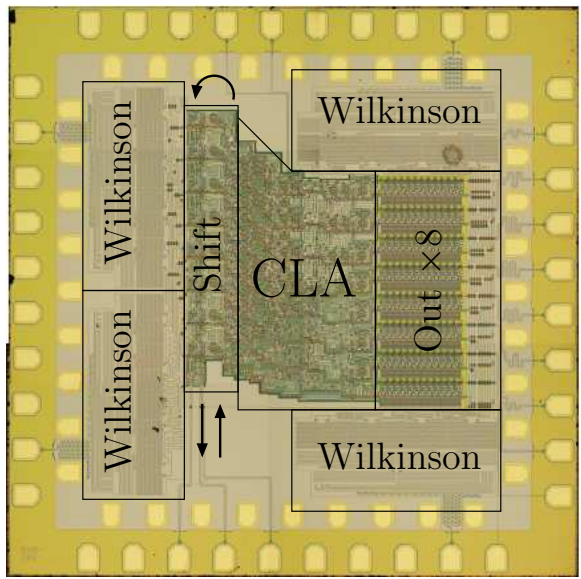}
\caption{Picture of the fabricated 8-bit CLA on a 5\,mm die, using a 2\,$\umu$m superconducting Nb process.
\label{fig3}}
\end{figure}

The CLA 5\,mm $\times$ 5\,mm chip is shown in Fig.~\ref{fig3}. The
circuit includes four passive, microwave Wilkinson power splitters for
clock distribution \cite{Oberg11WPS} and eight distributed amplifiers
\cite{Herr2010high} to output the result to room temperature. A
16-stage shift register with a single serial input is used to generate
the two input words. Taps on the shift register feed the CLA inputs:
starting with the LSB of input word A and working up, then wrapping
around to the MSB of word B and working back down. In this way an
arbitrary combination of input words can be applied every 16 clock
cycles, with the same bits reused in different combinations in the
intermediate clock cycles.

The circuit was designed for a Nb superconductor foundry service
\cite{Hypres} that is more than adequate to yield the
circuit. However, process integration scale is quite modest by CMOS
standards, with only four metal layers and 4\,$\umu$m wire pitch, which
limits circuit density. Two metal layers were used either for double
ground planes surrounding the logic gates or for clock stripline,
leaving only the remaining two metal layers for gates and
interconnect.

To accommodate physical layout, an additional idle phase consisting
only of delay cells was added before the last CLA column to allow the
three longest active interconnects to cover distance. This increased
the total number of clock phases to six, which amounts to 1.5 clock
cycles or 150\,ps latency at a 10\,GHz clock rate. In the final design
the three long active interconnects were replaced by 5\,$\Omega$ passive
superconducting striplines, which propagate signals at a
100\,$\umu$m/ps speed-of-light. The signals are received by active
interconnect circuits with the timing constraint that the receiver
must be near the peak of the clock waveform to receive the positive
pulse.

The clock distribution network is implemented as two eight-way
Wilkinson power splitters for the in-phase and quadrature clock
phases, and two identical Wilkinson devices used in reverse to
recombine the eight lines onto a single line \cite{Oberg11WPS}. In
this way the AC clock enters and exits the chip without ever
contacting chip ground. The 32\,$\Omega$ impedance of the clock lines
on chip is the largest impedance achievable using the bottom metal
layer, with a 2\,$\umu$m minimum line width, and using the top metal
layer as ground. The Wilkinson splitters were designed for a center
frequency of 7.5\,GHz, and optimized to give better than 30\,dB return
loss from 5-10\,GHz, and isolation better than 15\,dB over the same
range. A second version of the design used splitters centered at
15\,GHz. A six-section design was chosen to transform the 50\,$\Omega$
impedance of the feed line to the effective 4\,$\Omega$ impedance of
the eight 32\,$\Omega$ clock lines in parallel. The circuit design
requires a nominal clock amplitude of 2\,mA per line. Total input on
each of the two clock lines amounts to only 3.2\,mA rms.

\section{Circuit Test}

\begin{figure}[tbp]
\scalebox{0.40}{
\includegraphics{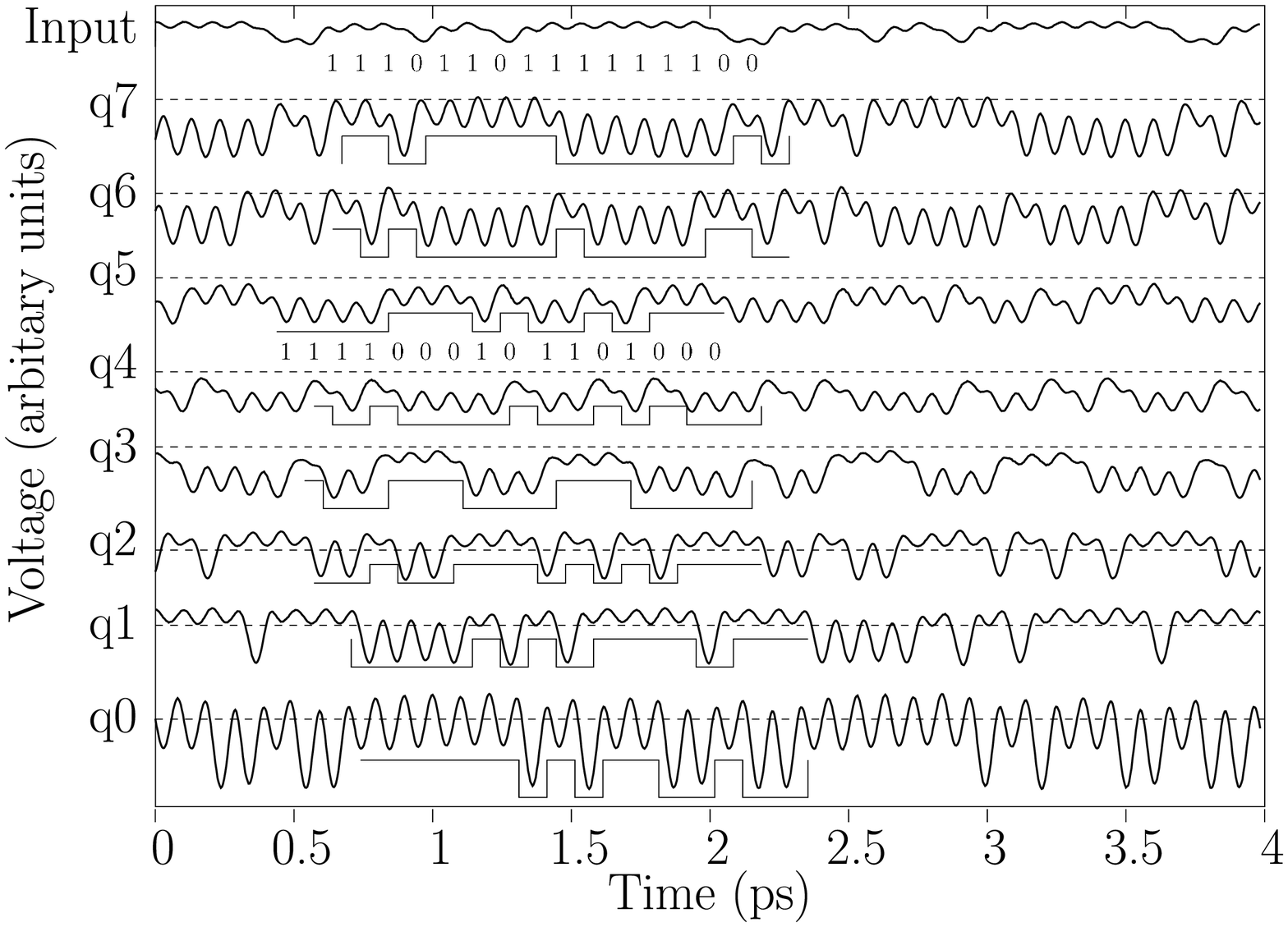}
}
\caption{The single-channel input and eight output waveforms from the
  adder as captured on a sampling oscilloscope at a clock rate of
  9.8\,GHz. The patterns are 16 bits long, and the addends consist of
  cyclic permutations of the serial input. One addend corresponds to
  8 serial input bits, and the other addend corresponds to the next 8
  bits in reverse order.
\label{fig4}}
\end{figure}

Each chip is mounted in a pressure-contact cryoprobe and cooled to
4.2\,K in a liquid helium dewar. The input is supplied from a digital
pattern generator through an inductive coupling to produce SFQ signals
on-chip, and returns to room temperature to be observed. The pattern
generator is phase-locked to a synthesizer that generates the clock
sinusoid, followed by a 90-degree hybrid to split the clock into
in-phase and quadrature. The clock lines also return to room
temperature after inductive couplings on chip and were terminated in
50\,$\Omega$ loads. The eight output bits from the adder are converted
from SFQ signals to source-terminated voltages of 2\,mV peak-to-peak
on-chip.

The high-speed input and output waveforms indicating correct digital operation of
the circuit are shown in Fig.~\ref{fig4}.  The input waveform is
non-return-to-zero, while the output waveforms are return-to-zero and
are inverted by low-noise amplifiers at room temperature.  The visible
feed through of the clock sinusoids to the outputs is frequency
dependent and is attributed to pickup in the cryoprobe package. Feed
through from the $-$2\,dBm clock lines to the outputs is down by
45\,dB, which is small in absolute terms but is comparable to the
$-$47\,dBm output levels. This effect could be eliminated in a future
chip design by using differential output.

\begin{figure}[tbp]
\scalebox{0.71}{
\includegraphics{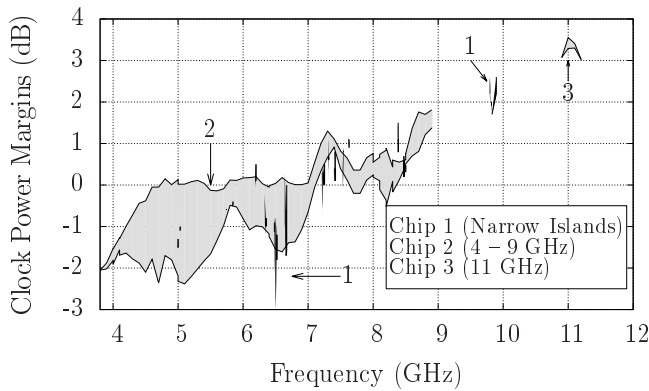}
}
\caption{Measured operating margins on clock power as a function of
  clock rate, based on observation of the most-significant output
  bit. Three chip types were tested that differed in the design of the
  Wilkinson power splitter and cryoprobe: 7.5\,GHz Wilkinson and
  narrowband cryoprobe (Chip 1), 7.5\,GHz Wilkinson and wideband
  cryoprobe (Chip 2), and 15\,GHz Wilkinson and wideband cryoprobe
  (Chip 3).
\label{fig5}}
\end{figure}

In simulation, the frequency-dependent operating margins on clock
power were limited by signal propagation time in the digital gates.
In test, power margins can be dominated by the microwave design of the
circuit, including the cryoprobe package. Different versions of the
chip were tested in two different cryoprobes, as shown in
Fig.~\ref{fig5}. The narrowband probe had a two-layer printed circuit
board (PCB) for signal traces and ground.  Chip test using this probe
showed very strong frequency dependence and was functional only in
narrow frequency bands in the range 5-10\,GHz. The highest observed
frequency for correct operation was 9.8\,GHz.  The wideband cryoprobe
had a redesigned PCB and used a second ground plane to minimize
crosstalk from the clock lines.  The probe transition from coaxial
line to stripline on-chip was optimized to give a return loss of
better than 20\,dB from dc up to 20\,GHz. Chip test using the
wideband probe produced a single large, continuous operating region
that extended from 4-9\,GHz. Another test, using the wideband probe
and a chip with the Wilkinson power network scaled up by a factor of
two in frequency, produced a small operating region observed around
11\,GHz.

\begin{figure}[tbp]
\includegraphics{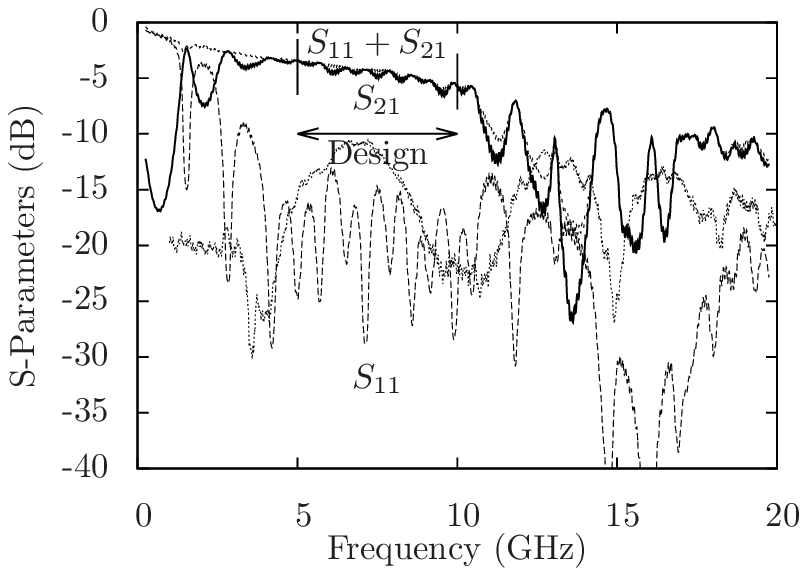}
\caption{S-parameters of the clock line for the 7.5\,GHz-Wilkinson
  power-splitter chip and the high-bandwidth probe, with the
  calibration plane at the probe head.
\label{fig6}}
\end{figure}

Digital operation of the chip (Chip~2 in Fig.~\ref{fig5}) can be
compared to S-parameter measurements of the clock distribution network
of the same chip, shown in Fig.~\ref{fig6}. The sum of transmitted and
reflected power measured on the clock line roughly corresponds to the
round-trip attenuation of the cables in the probe, indicating small
losses due to radiation and coupling to other lines across the entire
frequency range up to 20\,GHz. Transmitted power shows the bandwidth
of the Wilkinson splitter to be about 4-10\,GHz, and the center
appears to be consistent with the 7.5\,GHz design value. Small
reflected power is a key figure of merit, as resonances create
standing waves, leading to non-uniform biasing of the circuitry and
narrowed clock power operating margins.  Within the bandwidth of the
Wilkinson, reflected power is 10-22\,dB less than transmitted power,
with several resonances visible. Qualitatively similar behavior was
observed for the low bandwidth probe. However, there appears to be
little correlation between power margin measurements of the digital
circuit and these resonances. This indicates that S-parameter
measurements alone cannot adequately predict frequency-dependent
digital operation of the chip. In any case, for any specific
application the clock will be single-tone and thus narrow-band. This
significantly relaxes microwave design constraints.

\section{Power Dissipation Measurement}

A final test measured dynamic power dissipation of the circuit. This
is done by observing the attenuation of the clock signals due to the
power draw of device-switching events in the circuit. With an applied
clock power of about 0.6\,mW, the measurement must be sensitive to
better than one part per thousand to be able to resolve the expected
dissipation.

To achieve this accuracy we use a modulation technique where the input
data pattern is periodically chopped, alternating between all-zeros
and a pseudo-random pattern. Since none of the Josephson junction
devices in the circuit are active for an all-zeros input, this
modulates the AC clock signal at the data chopping frequency,
producing sidebands that are measurable with a spectrum
analyzer. Dynamic power dissipation of the circuit is calculated from
the power in the sidebands.

\begin{figure}[tbp]
\includegraphics{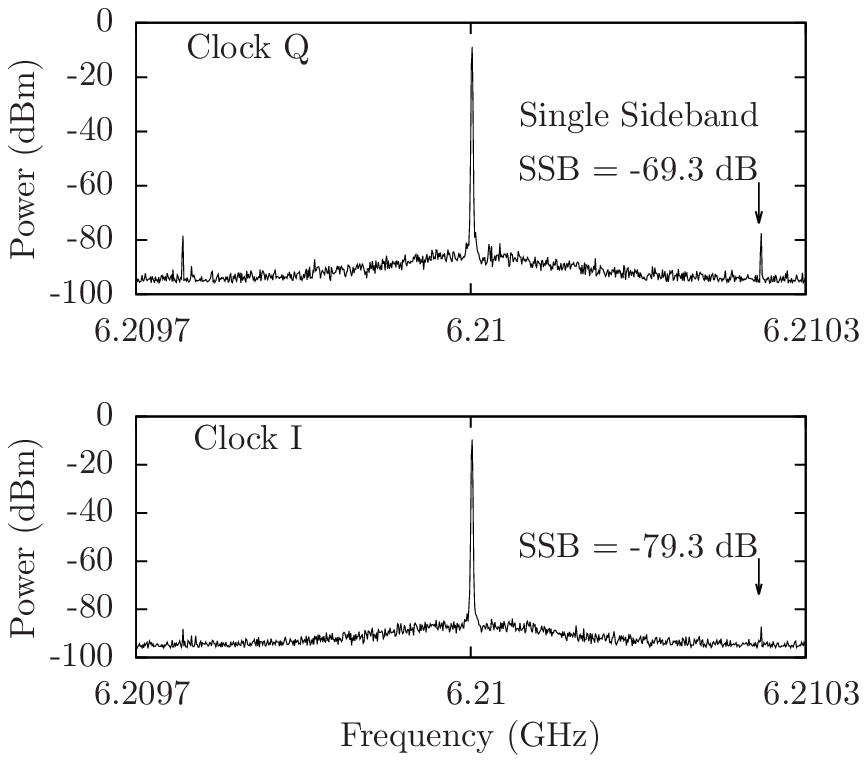}
\caption{Spectrum of the modulated clock output at 6.2\,GHz, measured
  on a spectrum analyzer, when the adder circuit was fed with a
  repetitive data input pattern of 12k pseudo-random bits followed by
  12k zero bits. The measured power ratio of the sidebands that are
  259\,kHz away from the fundamental is shown.
\label{fig7}}
\end{figure}

Fig.~\ref{fig7} shows the spectra of the two clock lines returned from
the chip. The two clock quadratures I and Q had respective power
levels of $-$2.4\,dBm and $-$2.0\,dBm at the chip, measured as the
geometric mean of applied and returned power to account for
attenuation in the probe cables. Sidebands are visible above and below
the carrier with a spacing of 259\,kHz, which corresponds to the
fundamental of the modulating square wave. The power ratio between a
single sideband and the carrier, $\textrm{SSB}=P_\textrm{SSB}/P_0$,
was measured to be $-$69.3\,dB for clock Q and $-$79.3\,dB for clock
I. Additional sidebands corresponding to higher odd harmonics of the
square wave fall outside the frequency range of our measurement.

The sidebands can arise either by amplitude modulation (AM), or by
phase modulation (PM). Both may be present in the the CLA adder circuit,
but only AM is indicative of power dissipation. The Josephson junction
circuit elements, which are inductively coupled to the clock, are
effectively in series with the clock line. The non-switching junction
can be modeled as an inductor, the switching junction as a
resistor. The switching junctions take power from the clock line,
producing AM, and decrease the propagation speed of the clock, which
gives rise to PM. We will first establish an upper bound on power
dissipation assuming purely AM, and then estimate dissipation
accounting for the relative contributions of AM and PM, using
previously reported measurements of an RQL circuit~\cite{RQL11}.

For the case of purely AM, the ratio of power dissipated to power in
the clock line, $\Delta P/P_0=(V^2_\textrm{hi}-V^2_\textrm{lo})/V^2_0$,
where $V_\textrm{hi}=V_0(1+2V_\textrm{sqr}/V_0)$ and
$V_\textrm{lo}=V_0(1-2V_\textrm{sqr}/V_0)$ are the maximum and minimum
amplitude of the clock sinusoid. $V_0$ is the amplitude of the clock
carrier and $V_\textrm{sqr}$ is the amplitude of the square wave
modulation. The factors of two account for the presence of double
sidebands, one above and one below the carrier frequency. The
fundamental of the square wave has an amplitude $4/\pi$ relative to
the amplitude of the square wave itself, so the amplitude of the
single sideband, $V_\textrm{SSB}=(4/\pi)V_\textrm{sqr}$. The normalized
power dissipation is
\begin{equation}
  \frac{\Delta P}{P_0}=2\pi\frac{V_\textrm{SSB}}{V_0}=
  2\pi\sqrt{\frac{P_\textrm{SSB}}{P_0}}.
\label{eqn1}
\end{equation}
This is an upper bound on power dissipation on the chip. It would also
be possible to produce the power spectrum measured in Fig.~\ref{fig7}
by phase modulation, without any power dissipation. Our previous
measurement of clock stablilty in an RQL shift register provides an
estimate of the relative contributions of AM and PM to the power
spectrum.

In order to estimate the AM and PM contributions to measured sideband
power we calculate the factors $m_a$ and $m_p$ corresponding to the
AM and PM amplitudes for a simple sine-modulated waveform
\begin{equation}
  V = V_0\,[1 + m_a \sin(\omega_m t)]\,\sin[\omega_c t + m_p \sin(\omega_m t)],
\label{eqn2}
\end{equation}
where $V_0$ and $\omega_c$ are the amplitude and frequency of the
carrier, and $\omega_m$ is the frequency of the modulation. From this
equation $m_a$ can be found by solving
$P_\textrm{lo}/P_\textrm{hi}=V^2_\textrm{lo}/V^2_\textrm{hi}=(1-2m_a)/(1+2m_a)$,
as the waveform amplitude ranges over $V_0(1\pm m_a)$. Similarly,
$m_p$ can be found by solving $\Delta
t=t_\textrm{hi}-t_\textrm{lo}=2m_p/\omega_c$, as the waveform phase
ranges over $\omega_c t \pm m_p$.  Both the data-modulated power ratio,
$P_\textrm{lo}/P_\textrm{hi}=0.91$, and a data-modulated time delay,
$\Delta t=1.4$\,ps, were previously reported in~\cite{RQL11} for an
RQL shift register operating at 6\,GHz. These measurements give
$m_a=0.023$ and $m_p=0.026$, indicating that AM and PM are about
equal at the clock rate of interest.

The amplitudes of the AM and PM sidebands add in quadrature, so for
equal AM and PM, only half of the power in the observed sideband is
attributable to AM. Including a correction factor of $1/2$ (or
equivalently, $-$3\,dB) to the AM sideband power of
equation~\ref{eqn1}, the normalized power dissipation is
\begin{equation}
  \frac{\Delta P}{P_0}=2\pi\sqrt{\frac{P_\textrm{SSB}/2}{P_0}}
  =8+\frac{1}{2}(\textrm{SSB}-3)\ \ \textrm{(in dB),}
\label{eqn3}
\end{equation}
where SSB is the power ratio of a single sideband to the carrier
expressed in dB. The correction factor reduces the estimate of power
dissipation compared to the upper bound set by purely AM by 30\%.

Using the measured values for SSB in equation~\ref{eqn3}, we calculate
that the active power dissipation was 970\,nW in clock Q and 280\,nW
in clock I. The difference in power dissipation between the two clock
quadratures is due to the output amplifiers, which are powered
exclusively on clock Q and make up 50\% of the junction critical
current on the chip. Using the measured values for $P_0$ on the two
clock lines, total power dissipation on chip amounts to
1.25\,$\umu$W. Excluding the amplifiers and the input shift register,
the CLA core makes up 42\% of total device critical current, so power
dissipation in the CLA amounts to 510\,nW.

This result is in agreement with the previously reported dynamic power
dissipation $P=0.33\,I_c\Phi_0Nf$ for a simple RQL shift
register. The CLA core, excluding the input shift register and output
amplifiers, has $N=815$ junctions of average critical current
$I_c=162\,\umu\mathrm{A}$, so power dissipation at $f=6.21$\,GHz is
expected to be 560\,nW. We note that this is equivalent to the static
power dissipation of a single bias resistor in the incumbent
superconductor logic family, RSFQ \cite{Likharev_91}, which typically
supplies 200\,$\umu$A from a 2.6\,mV bus.

\section{Conclusion}

We report an 8-bit carry look-ahead adder that advances reciprocal
quantum logic from the first benchmark experiments to a complex
circuit performing a recognizable logic function. The circuit confirms
the claims made for RQL for high speed and power-efficiency. The
circuit combines superconductor devices with design features that are
similar to CMOS, including combinational logic with fanout of four and
non-local interconnect. The RQL adder is high speed, with latency of
only 150\,ps at a clock rate of 10\,GHz. The design scales well. A
64-bit adder would have a latency of only 2 clock cycles, or 100\,ps
latency at 20\,GHz in a more advanced fabrication process. Based on
stability of the AC power, which also serves as a clock, the
integration scale of the circuit could be increased 100 times without
increasing the 1.2\,mW input power. Measured power dissipation in the
8-bit CLA adder was only 510\,nW at 6.2\,GHz. Bit energy is
approaching 1000\,$k_\textrm{B}\textrm{T}$, which is 100 to 1,000
times lower than high-performance CMOS at the 22\,nm node.

\begin{acknowledgments}
The authors thank Donald Miller for assistance with the power
measurement.
\end{acknowledgments}



\begin{thebibliography}{10}

\bibitem{RQL11}
Q.P. Herr, A.Y. Herr, O.T. Oberg, and A.G. Ioannidis,
{\em Ultra-Low-Power Superconductor Logic,}
J. Appl. Phys., 109 (2011), pp.~103903--103911.

\bibitem{Coolers}
R. Radebaugh,
{\em Cryocoolers: the state of the art and recent developments,}
J. Condens. Matter, 21 (2009), pp.~164219--164228.

\bibitem{Likharev_91}
K.K. Likharev and V.K. Semenov,
{\em {RSFQ} logic/memory family: a new {J}osephson-{J}unction digital
technology for sub-terahertz-clock-frequency digital systems,}
IEEE Trans. Appl. Supercond., 1 (1991), pp.~3--28.

\bibitem{e-RSFQ}
D.E. Kirichenko, S. Sarwana, and A.F. Kirichenko,
{\em Zero static power dissipation biasing of {RSFQ} circuits,}
IEEE Trans. Appl. Supercond., 21 (2011), pp.~776--779.

\bibitem{kameda1999self}
Y. Kameda, S.V. Polonsky, M. Maezawa, and T. Nanya
{\em Self-timed parallel adders based on {DI RSFQ} primitives,}
IEEE Trans. Appl. Supercond., 9 (1999), pp.~4040--4045.

\bibitem{bunyk_99}
P. Bunyk and P. Litskevitch,
{\em Case study in {RSFQ} design: fast pipelined parallel adder,}
IEEE Trans. Appl. Supercond., 9 (1999), pp.~3714--3720.

\bibitem{Filippov11}
T.V. Filippov, A. Sahu, A.F. Kirichenko, I.V. Vernik,
M. Dorojevets, C.L. Ayala, and O.A. Mukhanov,
{\em 20 {GH}z operation of an asynchronous wave-pipelined {RSFQ}
arithmetic-logic unit,}
Physics Procedia, 36 (2012), pp.~59--65.

\bibitem{loss}
A. Vayonakis, C. Luo, H.G. Leduc, R. Schoelkopf and J. Zmuidzinas,
{\em The millimeter-wave properties of superconducting microstrip lines,}
AIP Conf. Proc., 605 2002), pp.~539--542.

\bibitem{wrspice}
S.R. Whiteley,
{\em Josephson junctions in {SPICE}3,}
IEEE Trans. Magn., 27 (1991), pp.~2902--2905.

\bibitem{Oberg11WPS}
O.T. Oberg, Q.P. Herr, A.G. Ioannidis, and A.Y. Herr,
{\em Integrated Power Divider for Superconducting Digital Circuits,}
IEEE Trans. Appl. Supercond., 21 (2011), pp.~571--574.

\bibitem{Herr2010high}
Q.P. Herr,
{\em A high-efficiency superconductor distributed amplifier,}
Supercond. Sci. Technol., 23 (2010), pp.~022004--022008.

\bibitem{Hypres}
{\em Hypres {N}b design rules,}
www.hypres.com, 2010.

\end{thebibliography}
\end{document}